\documentclass{aa}
\usepackage{upgreek}
\usepackage{graphicx}
\usepackage[varg]{txfonts}
\usepackage{natbib}
\usepackage{soul}
\newlength{\linwx}
\usepackage{color}

\begin{document}


\title{Inner rocky super-Earth formation: distinguishing the formation pathways in viscously heated and passive discs}

\author{
Bertram Bitsch 
}
\offprints{B. Bitsch,\\ \email{bitsch@mpia.de}}
\institute{
Max-Planck-Institut f\"ur Astronomie, K\"onigstuhl 17, 69117 Heidelberg, Germany
}
\abstract{Observations have revealed that super-Earths (planets up to 10 Earth masses) are the most abundant type of planets in the inner systems. Their formation is strongly linked to the structure of the protoplanetary disc, which determines growth and migration. In the pebble accretion scenario, planets grow to the pebble isolation mass, at which the planet carves a small gap in the gas disc halting the pebble flux and thus its growth. The pebble isolation mass scales with the disc's aspect ratio, which directly depends on the heating source of the protoplanetary disc. I compare the growth of super-Earths in viscously heated discs, where viscous heating dissipates within the first million years, and discs purely heated by the central star with super-Earth observations from the Kepler mission. This allows two formation pathways of super-Earths to be distinguished in the inner systems within this framework. Planets growing within 1 Myr in the viscously heated inner disc reach pebble isolation masses that correspond directly to the inferred masses of the Kepler observations for systems that feature planets in resonance or not in resonance. However, to explain the period ratio distribution of Kepler planets -- where most Kepler planet pairs are not in mean motion resonance configurations -- a fraction of these resonant chains has to be broken. In case the planets are born in a viscously heated disc, these resonant chains thus have to be broken without planetary mergers, for example through the magnetic rebound effect, and the final system architecture should feature low mutual inclinations. If super-Earths form either late or in purely passive discs, the pebble isolation mass is too small (around 2-3 Earth masses) to explain the Kepler observations, implying that planetary mergers have to play a significant role in determining the final system architecture. Resonant planetary systems thus have to experience mergers already during the gas disc phase, so the planets can get trapped in resonance after reaching 5-10 Earth masses. In case instabilities are dominating the system architecture, the systems should not be flat, but feature mutually inclined orbits. This implies that future observations of planetary systems with radial velocities (RV) and transits (for example through the Transiting Exoplanet Survey Satellite (TESS) and its follow up RV surveys) could distinguish between these two formation channels of super-Earth and thus constrain planet formation theories.
}
\authorrunning{Bitsch}\titlerunning{Inner rocky super-Earth formation}\maketitle

\section{Introduction}
\label{sec:Introduction}

Recent observations have revealed that close-in super-Earths are very abundant and even more than 50\% of all stars could host super-Earths \citep{2011arXiv1109.2497M, 2018AJ....156...24M, 2018arXiv180502660Z}. These super-Earth planets are typical of a few Earth masses and super-Earths within the same system are typical of similar size \citep{2018AJ....155...48W} and thus presumably mass, within a factor of two.

Additionally, observations of protoplanetary discs in different stellar cluster environments have shown that the building blocks of planetesimals and planets (millimeter-centimeter-sized dust grains) are presumably not very abundant as soon as the protoplanetary discs become older than 1 Myr \citep{2016ApJ...828...46A, 2017AJ....153..240A, 2019ApJ...875L...9W}, posing a problem for planet formation \citep{2018A&A...618L...3M}. This either implies that there is a hidden reservoir of building blocks (e.g. in the form of planetesimals or small dust that still needs to grow), that the dust masses are underestimated due to the optical thickness of the disc \citep{2019arXiv190402127Z}, or that planet formation mainly happens within the first million years after star formation, as is also assumed to explain the observed rings and gaps in protoplanetary discs \citep{2018ApJ...869L..47Z, 2019MNRAS.tmp.1807N}.

Planet formation theories for super-Earths have to match these constraints. If super-Earths form from the accretion of planetesimals in high density protoplanetary discs, the formation time scales are very short as soon as planetesimals become available \citep{2015A&A...578A..36O}. However, the resulting systems do not necessarily harbour equal mass planets but can instead feature a strong radial mass gradient \citep{2015A&A...578A..36O}. Additionally the resulting systems are too tightly packed compared to the observations. This makes the formation of these systems via planetesimal accretion probably difficult. However, \citet{2018A&A...615A..63O} have shown that systems with roughly equal mass planets can form via planetesimal accretion in discs, which evolve primarily due to disc winds \citep{2016arXiv160900437S}.

Another mechanism to form close-in super-Earth planets is by the accretion of pebbles \citep{2010MNRAS.404..475J, 2010A&A...520A..43O, 2012A&A...544A..32L}. In this scenario a high initial local density of solids is not needed, because the small millimeter-centimeter-sized pebbles drift through the disc \citep{1977MNRAS.180...57W, 2008A&A...480..859B} from far out and can then be accreted by planetary embryos in the inner disc. This implies that a much lower initial disc surface density can be assumed and super-high densities as in the proposed minimum-mass-extrasolar-nebular, MMEN, \citep{2013MNRAS.431.3444C} are not needed for planet formation. The MMEN has also been criticised by \citet{2014MNRAS.440L..11R}, because the diversity of super-Earth systems can not be explained within this simple power-law disc model for in-situ planet formation, which is the prefered planet formation mode used to derive the MMEN. Additionally, the timescales of pebble accretion within the inner disc are short enough, so that formation of full grown super-Earths can easily happen within 1 Myr \citep{2019arXiv190208694L, 2019arXiv190208772I}.

Planetary growth in the pebble accretion scenario continues until the pebble isolation mass \citep{2012A&A...546A..18M, 2014A&A...572A..35L, 2018arXiv180102341B, 2018A&A...615A.110A} is reached. At this mass, the planetary embryo opens a small gap in the protoplanetary disc, which inverts the pressure gradient of the protoplanetary disc exterior to the planet, resulting in the accumulation of pebbles exterior to the planet \citep{2006A&A...453.1129P}. The planet has thus shielded itself from the pebbles and stops growing via pebble accretion.

The pebble isolation mass in itself is a function of the disc's thermal parameters. It depends mainly on the disc's aspect ratio $H/r$, and weakly on the radial pressure gradient $\partial\ln P / \partial \ln r$, and the disc's viscosity \citep{2018arXiv180102341B}. It scales as
\begin{equation}
\label{eq:MisowD}
  M_{\rm iso} = 25 f_{\rm fit} {\rm M}_{\rm E} + \frac{\Pi_{\rm crit}}{\lambda} {\rm M}_{\rm E} \ ,
\end{equation}
with $\lambda \approx 0.00476 / f_{\rm fit}$, $\Pi_{\rm crit} = \frac{\alpha_{\rm disc}}{2\tau_{\rm f}}$, and
\begin{equation}
\label{eq:ffit}
 f_{\rm fit} = \left[\frac{H/r}{0.05}\right]^3 \left[ 0.34 \left(\frac{\log(\alpha_3)}{\log(\alpha_{\rm disc})}\right)^4 + 0.66 \right] \left[1-\frac{\frac{\partial\ln P}{\partial\ln r } +2.5}{6} \right] \ ,
\end{equation}
where $\alpha_{\rm disc}$ is the $\alpha$ parameterization of the disc's viscosity, $\alpha_3 = 0.001$, and $\tau_{\rm f}$ is the particles Stokes number. The Stokes number in the used model model varies roughly from 0.05 to 0.2; see appendix~\ref{ap:growth} for more details. This corresponds to the typical dominant Stokes number of particle growth limited by radial drift \citep{2012A&A...539A.148B, 2014A&A...572A.107L, 2018A&A...609C...2B}.

The consequence of the pebble isolation mass is that the final planetary mass in the pebble accretion scenario is set by the disc structure and does not scale with the initial distribution of solids in the disc as long as planetery mergers do not contribute to the final planetary mass. This idea is supported by the recent analysis of the planetary masses of the Kepler observations which has lead to the speculation that planets grow to the thermal mass of the protoplanetary disc\footnote{The disc's thermal mass is defined as the planet-to-star mass ratio at which its Hill radius exceeds the gas scale height. The pebble isolation mass is caused by a gap in the pebble distribution, which is achieved at lower planetary masses (e.g. \citealt{2006A&A...453.1129P}), resulting in the fact that the pebble isolation mass is lower than the disc's thermal mass, even though the scaling with the disc's aspect ratio is similar. This also allows \citet{2019ApJ...874...91W} to have a match to the Kepler observations with the thermal mass, even though the used disc model features very low temperatures, see section~\ref{sec:disc}.} \citep{2019ApJ...874...91W}. This implies that the final planetary mass could be determined by the disc structure in itself and not by the amount of available solids. 

I compare the pebble isolation masses of planets growing in the inner systems of discs dominated by viscous or stellar heating. Depending on the disc structure I distinguish between two formation pathways of super-Earth systems in the inner protoplanetary disc. One pathway is dominated by collisions and planetary mergers \citep{2017MNRAS.470.1750I, 2019arXiv190208772I, 2019arXiv190208694L}, while the other formation pathways requires only a minimal amount of collisions and instabilities, because the planets reach their final masses completely by the accretion of pebbles during the gas disc phase, which corresponds to the pebble isolation mass.

I focus here strictly on the formation of super-Earths within the inner 2-3 AU of protoplanetary discs, where the building blocks are presumably rocky due to their location interior to the water ice line. In principle super-Earths could form in the outer disc, where the pebble isolation mass remains large for a long time, and then migrate inwards \citep{2017MNRAS.470.1750I, 2019arXiv190208772I}. However, these bodies would all be water ice rich and could thus not form the population of rocky super-Earths \citep{2019arXiv190208772I}, unless the bodies are very efficient at losing water ice.

This paper is structured as follows. In section~\ref{sec:disc} I describe the protoplanetary disc models. I then discuss the formation of super-Earths in section~\ref{sec:formation} and describe the two formation pathways of inner rocky super-Earths in section~\ref{sec:chanel}. I then discuss the model assumptions in section~\ref{sec:discussion} and finally summarise in section~\ref{sec:summary}.

\section{Disc model}
\label{sec:disc}

The evolution of the protoplanetary disc was long thought to be driven by viscous evolution \citep{1974MNRAS.168..603L}. However, recent more detailed simulations have revealed that the transport of angular momentum can be carried by disc winds \citep{2013ApJ...769...76B, 2013ApJ...772...96B, 2016arXiv160900437S}, but viscous disc evolution could still be consistent with observations \citep{2018MNRAS.474...88H}.

Important for the here presented scenario is mostly the disc's thermal structure. For this I relied mostly on the disc evolution model of \citet{2015A&A...575A..28B}, which is computed from 2D radiation hydrodynamic simulations including viscous and stellar heating as well as radiative cooling. The time evolution of the disc is then related to a decrease in the disc's accretion rate following observations \citep{1998ApJ...495..385H, 2009AIPC.1158....3M}.

As the disc evolves in time, it cools down and with it the disc's aspect ratio decreases, resulting in a decrease in the pebble isolation mass in time (Fig.~\ref{fig:Planets}). Additionally the position of the water ice line moves inwards in time and comes close to 1 AU after 1 Myr. This property is not unique to the here used disc model, but is a common outcome of disc evolution models that include viscous heating \citep{2011ApJ...738..141O, 2015arXiv150303352B}. The here used disc model is thus a representative of typical viscous disc models, making our results wider applicable.

Additionally a comparison to a passively heated disc model, as there has been a recent debate if viscous heating is actually active in the inner regions of protoplanetary discs \citep{2019ApJ...872...98M}, is discussed. The here used passive disc model follows an MMSN like structure in gas surface density
\begin{equation}
 \Sigma_{\rm g,pas} = 1700 \left(\frac{r}{\rm AU}\right)^{-3/2} \frac{\rm g}{\rm cm^2} \ .
\end{equation}

The aspect ratio of a passively heated disc follows a $2/7$ power law in orbital distance for the outer disc \citep{1997ApJ...490..368C, 2004A&A...417..159D, 2013A&A...549A.124B}. However, in the very inner regions of the protoplanetary disc, the sunlight grazing angle is dominated by the finite angular size of the star, which changes the disc profile. The passive disc profile used here follows for H/r the scaling relations of the disc model presented in \citet{2016A&A...591A..72I} for the outer disc and uses the scaling of \citet{2019ApJ...874...91W} for the inner disc\footnote{The disc profile of the outer disc in \citet{2019ApJ...874...91W} leads to a very low temperature. In that disc model, the water ice line is very close to 1 AU, which seems unrealistic to form the Earth in our own solar system dry. \citet{2019ApJ...874...91W} use the disc model of \citet{2010AREPS..38..493C}, based on \citet{1997ApJ...490..368C}, who use a ratio of 4 for the visible photosphere to the pressure scale height, even though a value of 6 is claimed to be used in \citet{2019ApJ...874...91W}, which would lead to higher temperatures. Other differences could arise from varying opacity prescriptions between the models.}.

The passive disc profile is given as
\begin{eqnarray}
\label{eq:disc}
 \left(\frac{H}{r}\right)_{\rm out} &=& 0.024 \left(\frac{R_{\rm \star}}{R_\odot}\right)^{2/7} \left(\frac{T_{\rm \star}}{T_\odot}\right)^{4/7} \left(\frac{r}{\rm AU}\right)^{2/7} \ , \quad r\ge r_{\rm trans} \\
 \left(\frac{H}{r}\right)_{\rm in} &=& 0.0173 \left(\frac{R_{\rm \star}}{R_\odot}\right)^{3/8} \left(\frac{T_{\rm \star}}{T_\odot}\right)^{1/2} \left(\frac{r}{\rm AU}\right)^{1/8} \ , \quad r\le r_{\rm trans} \nonumber
\end{eqnarray}
For the passively irradiated disc model, the stellar parameters are given as $R_\star=3.096R_\odot$ and $T_\star=4397$K, corresponding to an early solar type star \citep{2015A&A...577A..42B}. The viscous disc model, on the other hand, uses a star that is initially 1 Myr old with $R_\star=2.02R_\odot$ and $T_\star=4470$K \citep{2015A&A...575A..28B} that changes its luminosity in time following \citet{1998A&A...337..403B}. However, the change of stellar irradiation on a disc dominated by viscous heating is minimal \citep{2015A&A...575A..28B}, implying that the differences in stellar irradation are only important for the outer regions of the disc in this case which are dominated by stellar heating. The passive disc model is calculated for a younger star to give a maximum amout of heating. The transition between both regimes for the passive disc model (eq.~\ref{eq:disc}) is given for our parameters at $r_{\rm trans}\approx 0.22$ AU and the water ice line is located at $\approx$1.65 AU.

Additionally an exponential decay of the gas disc surface density with $\exp{(-t/t_{\rm disc})}$ and $t_{\rm disc}=3$ Myr is applied for the passive disc model. In this case, for simplicity, the temperature does not evolve in time by assuming that the stellar luminosity, which determines the disc's temperature \citep{1997ApJ...490..368C, 2004A&A...417..159D, 2013A&A...549A.124B}, does not evolve during the gas-disc phase. This decay, however, is taken into account in \citet{2015A&A...575A..28B}, which is why the pebble isolation mass in the outer regions of the protoplanetary disc decreases in time (see Fig.~\ref{fig:Planets}).

\section{Formation of inner super-Earths}
\label{sec:formation}

The masses of planets in the Kepler sample is mostly unknown, because only the radius of planets is determined by transits. In order to convert the radius of the planet to a planetary mass, the mass-radius relationship of \citet{2016ApJ...825...19W} derived from a sub-sample that contains also radial velocity (RV) data and thus planetary masses is used. This allows to determine the masses of planets from their radius via the equation
\begin{equation}
 M_{\rm P} = 2.7 \left( \frac{R_{\rm P}}{R_{\rm E}} \right)^{1.3} \rm M_{\rm E} \ ,
\end{equation}
where $R_{\rm P}$ and $M_{\rm P}$ correspond to the planetary radius and mass, respectively. This relationship is only valid for planets up to 4 Earth radii, where planets with larger radii are assumed to be planets dominated by gas rather than solid material. Nevertheless, planets above several Earth masses can start to accrete small gaseous envelopes, which influences the mass-radius relationship.

\begin{figure*}
 \centering
 \includegraphics[scale=1.3]{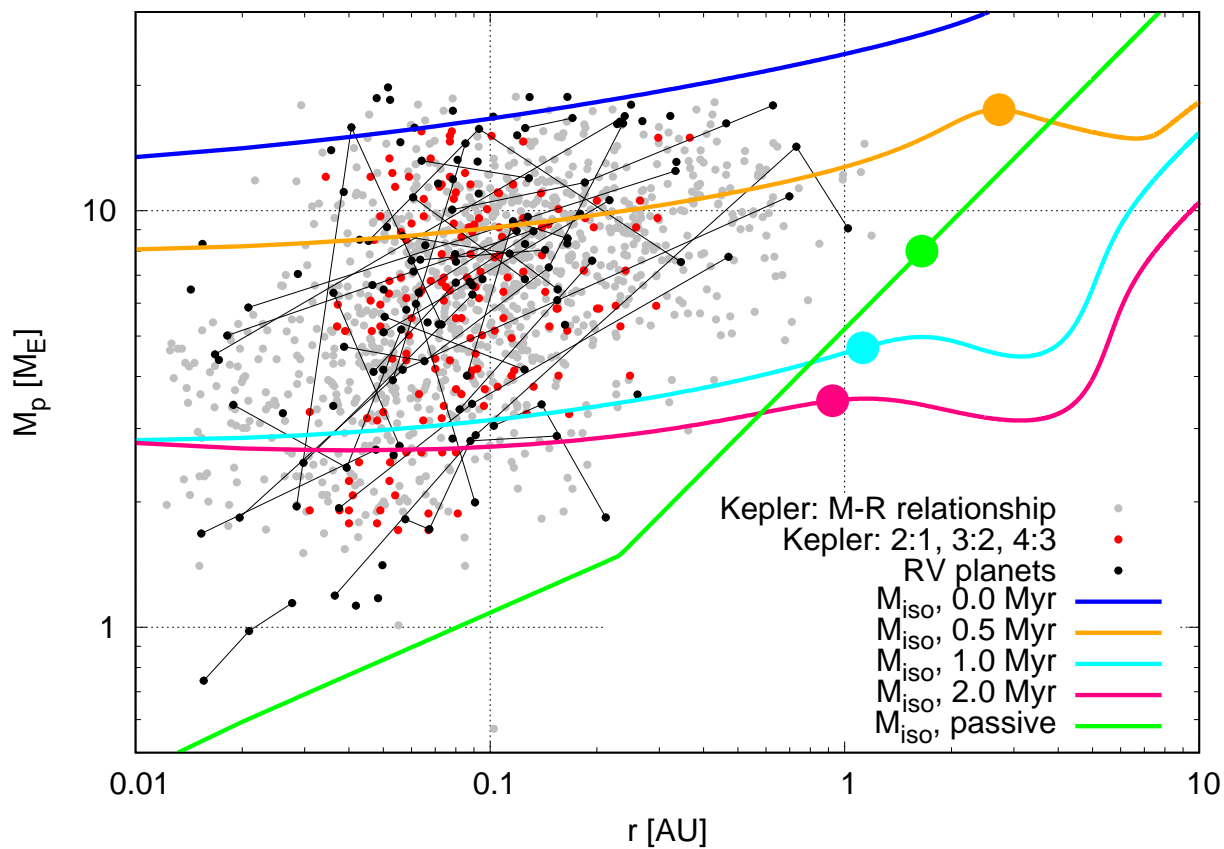} 
 \caption{Planetary masses of Kepler systems derived from the mass-radius relationship from \citet{2016ApJ...825...19W}, with cut-off at planetary radii of 4 $R_{\rm E}$ (grey dots) and super-Earths detected by RV (black dots) with cut-off at 20 Earth masses. The black lines connect planets within the planetary system (for RV systems only). The red dots correspond to Kepler planets in multiple systems with period ratios of $\sim$2:1, 3:2 and 4:3 to their neighbouring planets. Only planets orbiting around solar mass stars (0.8-1.2 ${\rm M}_\odot$) are shown. The colour lines correspond to the pebble isolation mass ($M_{\rm iso}$) from \citet{2018arXiv180102341B} with aspect ratios derived from 2D disc evolution models including viscous and stellar heating of \citet{2015A&A...575A..28B} as well as a passive disc (green). Just for plotting purposes a Stokes number of 0.1 is used, but the Stokes number of the pebbles varies within the simulations from 0.05 to 0.2. The coloured dots on the lines correspond to the position of the water ice line in these disc models. The disc evolution model implies that if planets from in the inner disc regions that their formation has to be completed within 1 Myr after the disc formation, because otherwise the pebble isolation mass becomes too small to explain the masses of the observed exoplanets. If planets form further out and migrate inwards, the formation time can be much later than 1 Myr, however these planets would form in the cold regions of the protoplanetary disc and would be predominantly of icy composition in contrast to observations that predict that close-in super-Earths should be predominantly rocky \citep{2013ApJ...775..105O, 2017ApJ...847...29O, 2018ApJ...853..163J}.  Late planet formation or planet formation in the passive disc implies that collisions between the bodies are needed to explain the observations.
   \label{fig:Planets}
   }
\end{figure*}

In Fig.~\ref{fig:Planets}, I show the masses of the planets from the Kepler data derived from the mass-radius relationship as well as the masses of planets detected via the RV method (black dots). The planets detected by Kepler that have period ratios of $\sim$2:1, 3:2, and 4:3 with their neighbouring planets are marked in red. Additionally I show the pebble isolation mass calculated with $\alpha=0.001$ for the disc model of \citet{2015A&A...575A..28B} and of the passive disc model using the same $\alpha$ value. The influence of different $\alpha$ parameters on $M_{\rm iso}$ are shown in appendix~\ref{ap:Miso}. The pebble isolation mass in these plots marks the upper limit to which planets can grow purely by pebble accretion -- without any mergers -- as a function of orbital distance. If a planet forms in a region of the disc where the pebble isolation mass is higher, it can still migrate inwards and be parked at the inner edge of the protoplanetary disc \citep{2015A&A...582A.112B}. Planets forming further out and migrating inwards can thus reach masses that are higher then the local pebble isolation mass in the inner disc.

However, as the disc evolves in time, the pebble isolation mass decreases and planets that form after 1 Myr within 1 AU do not reach pebble isolation masses high enough to explain the bulk of the observed planets without planetary mergers. If the growth of super-Earths via pebble accretion happens within 2-3 AU from the central star, they have to form within the first Myr after disc formation if they are purely formed via pebble accretion and without mergers between planetary embryos.

In principle collisions between planets after the gas-disc phase could increase their masses \citep{2019arXiv190208772I}. However the resulting systems would not feature any resonant configuration between the planets, making the formation of the resonant systems very difficult after gas disc dissipation.

\begin{figure}
 \centering
 \includegraphics[scale=0.7]{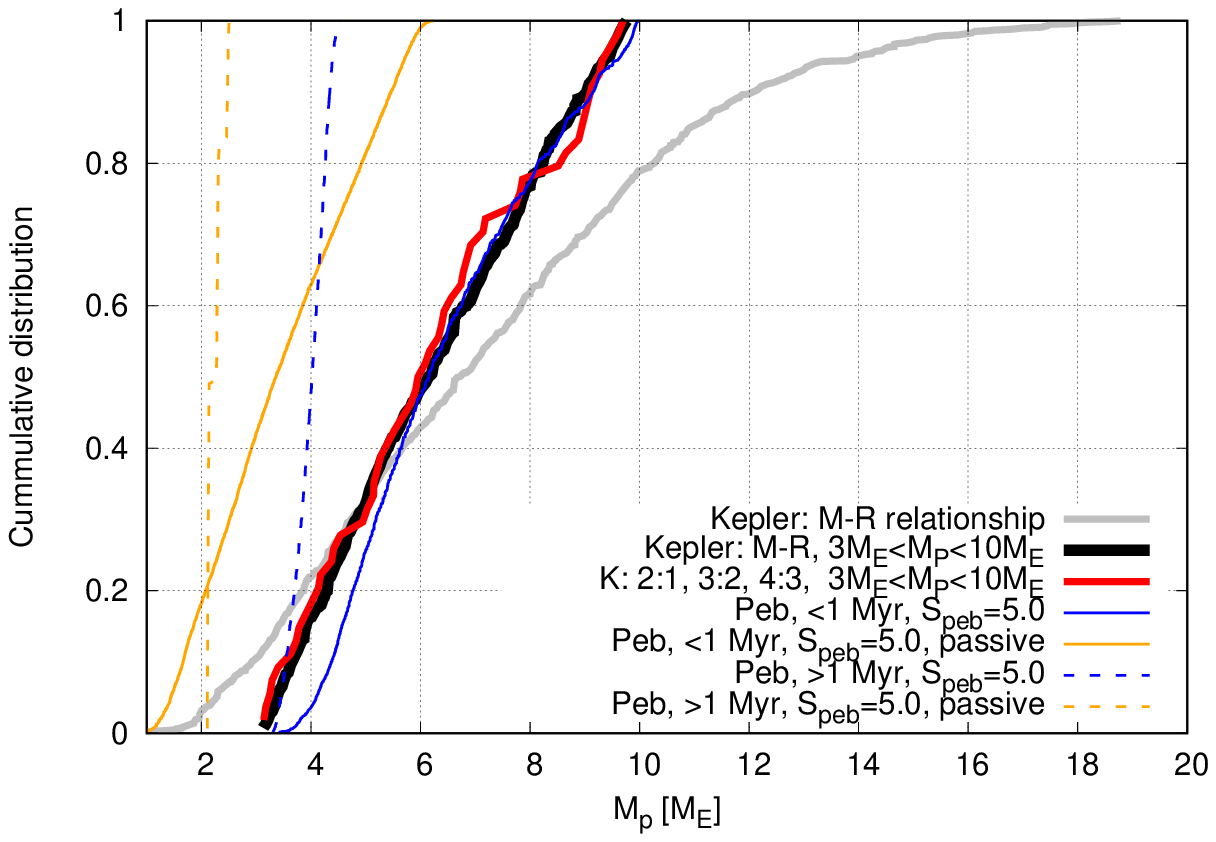} 
 \caption{Planetary masses for Kepler systems derived from the mass-radius relationship from \citet{2016ApJ...825...19W}, where the grey color shows the same sample as in Fig.~\ref{fig:Planets}. In black the Kepler systems with planetary masses ranging from 3 to 10 ${\rm M}_{\rm E}$ are displayed. In red the masses from systems in period ratios close to $~\sim$ 2:1, 3:2, and 4:3 in the same mass range are shown. The masses of the planets from the simulations, divided into planets that reach the pebble isolation mass before (solid lines) or after 1 Myr (dashed lines) are also shown. The planets growing in the passive disc model do not reach masses conform with the Kepler sample, because the pebble isolation mass is too low in the inner disc, implying that mergers between the planets have to play a key role in this case.
   \label{fig:Masses}
   }
\end{figure}

Using the same disc model and pebble evolution prescriptions as in \citet{2019arXiv190208772I}, the growth and migration (see appendix~\ref{ap:growth} for a detailed model description) of single planetary embryos with starting masses of 0.01 ${\rm M}_{\rm E}$ and an initial radial distribution of 0.2 to 3.0 AU is modelled. The final planetary masses of planets that have a final orbital position within 1 AU are shown in Fig.~\ref{fig:Masses}, where the simulations with $S_{\rm peb}=5.0$ are shown. I divide the results of the simulations in planets that have reached the pebble isolation mass within 1 Myr and in planets that reach the pebble isolation mass after 1 Myr.

I compare the simulations mainly to the sample of Kepler planets between 3 and 10 ${\rm M}_{\rm E}$ because larger planets probably start to accrete gas efficiently \citep{1996Icar..124...62P, 2014ApJ...786...21P} and the final formation of smaller planets than 3 ${\rm M}_{\rm E}$ is probably a result of giant impacts rather than pebble accretion \citep{2019arXiv190208694L}. The final planetary masses of the simulations match the Kepler observations very well, if the pebble flux is large enough (see appendix~\ref{ap:growth}). If the pebble flux is very low, planets do not grow very much by pebble accretion and would form Mars-size embryos \citep{2019arXiv190208694L}. For larger pebble fluxes the planets always reach pebble isolation mass, which is independent of the pebble flux and only depends on the protoplanetary disc structure.

A similar conclusion was also derived in \citet{2019ApJ...874...91W}, who state that Kepler planets grow to a typical mass, which is set by the disc's thermal mass. The disc's thermal mass scales with $(H/r)^3$, similar as the pebble isolation mass (eq.~\ref{eq:MisowD}). It is clearly shown here that planetary growth via pebble accretion in viscously heated discs within 1 Myr could naturally explains the mass distribution of observed super-Earths.

The formation of planets within the first Myr of the protoplanetary disc is also thought to have happened in our own solar system. \citet{2017LPI....48.1386K} concluded through meteoritic evidence that the reservoirs of non-carbonaceous meteorites and carbonaceous meteorites were spatially separated in the protoplanetary disc around the young sun at about $\approx$1 Myr. This separation can be achieved by a growing planet that stops the inward flux of particles, which would correspond to Jupiter's core exceeding the pebble isolation mass in the outer disc. In addition, observations of young (1-2 Myr) protoplanetary discs featuring rings in the millimeter-dust distribution are thought to be caused by growing planets (e.g. \citealt{2018ApJ...869L..47Z, 2019MNRAS.tmp.1807N}) of at least a couple of Earth masses. As orbital time-scales are even shorter in the inner disc, these lines of evidence argue for an early planet formation.

In contrast, planets formed in a purely passive disc are much smaller than planets formed in the viscously heated disc. This is related to the fact that the pebble isolation mass is much smaller (2-3 $\rm M_{\rm E}$) in the inner regions of the disc (Fig.~\ref{fig:Planets}). Collisions between planets can help to increase their masses to the mass estimates of super-Earths derived from the Kepler sample. Resonant planetary systems in this case have to undergo mergers during the gas-disc phase, so that the final more massive planets can be trapped in resonant configurations again.

\section{Two formation channels of super-Earth growth in the inner system}
\label{sec:chanel}

From the simulations and their comparison to the mass estimates from the Kepler mission, two formation channels emerge depending if the underlying protoplanetary discs are viscously heated or not. In the following I outline these two formation channels for inner rocky super-Earths growing by pebble accretion.

\subsection{Viscously heated disc}

In the viscously heated disc, the pebble isolation mass is initially high and planets reaching the pebble isolation mass can easily grow to masses of a few Earth masses within 0.5-1 Myr, if the pebble accretion rate onto the planet is high enough \citep{2019arXiv190208694L}, see Fig.~\ref{fig:Masses2}. These final masses match the masses of super-Earths estimated from the Kepler sample (Fig.~\ref{fig:Masses}). As the planets grow and migrate they pile up in resonant chains close to the inner edge, naturally explaining the resonant planetary systems, which follow a very similar mass distribution (red curve in Fig.~\ref{fig:Masses}). 

In the work by \citet{2017MNRAS.470.1750I} and \citet{2019arXiv190208772I}, these resonant chains can become unstable and thus create systems out of resonant configurations. However, if this happens, the chains of planets that form early in the disc have such high masses that the resulting planetary systems after planetary mergers are mostly too large compared to the Kepler sample \citep{2019arXiv190208772I}. This implies that another mechanism must be the dominant source of breaking the resonant chains of super-Earths with individual masses above 5-6 $\rm M_{\rm E}$. 

The resonant chains could also be broken by the retreat of the inner edge of the gas disc at dispersal \citep{2017A&A...601A..15L, 2017A&A...606A..66L}, which can move the planets out of the resonance.  Another possibility is that the lifetime of the protoplanetary disc is short enough that the planets do not migrate all the way to the inner edge of the disc and thus they do not form resonant chains in the first place \citep{2017AJ....153..222B}. However, this depends on the exact planet migration prescription and if planet traps exits besides the inner edge of the protoplanetary disc, e.g. at the water ice line \citep{2015A&A...575A..28B}.

In both cases, however, only a minimal amount of these non-resonant systems should become unstable after gas disc dissipation to not increase the masses of the super-Earths to values above the Kepler estimates. Additionally, the resulting planetary systems should feature relatively low mutual inclinations between the individual planets. This makes it difficult to explain the high number of single planet observations in the Kepler data, which can be explained by mutual inclinations between the individual planets caused by a high rate of instabilities \citep{2019arXiv190208772I}.

\subsection{Passive disc}

The case of a passive disc, where viscous heating is not present (which is compareable to planets forming in the viscous disc model at late times), the pebble isolation mass is smaller and the planetary masses are too small to match the Kepler mass estimates directly. Giant impacts between the bodies could increase the masses of these super-Earths. This is observed in the simulations of \citet{2019arXiv190208772I}, where long chains of low mass planets (2-3 $\rm M_{\rm E}$) that form in old discs become unstable and their final masses give a good match to the Kepler mass estimates. However, if the planetary systems become unstable after the gas disc dissipation, the resulting planetary systems will not be in a resonant configuration. Additionally, the number of planetary chains that become unstable does not allow a good match to the number of planets detected by transits \citep{2019arXiv190208772I}, where mostly single planets are detected by Kepler. The large fraction of single detections can be explained by multi-planet systems with large mutual inclinations, where \citet{2019arXiv190208772I} suggested that 95\% of the initially resonant planetary systems must have become unstable after the gas disc phase. However, the systems containing the smaller mass planets do not reach that level of instability in \citet{2019arXiv190208772I}. On the other hand, \citet{2019arXiv190208772I} are missing mechanisms that could drive further instabilities, e.g. tides and influences from leftover planetesimals which could increase the instability rate and allow a match of the simulations with the observations.

A way to produce massive super-Earths in resonant systems from smaller bodies are collisions during the gas phase. As the bodies collide, they grow, but as they are still embedded in the gas phase the planets can still migrate and be trapped in resonant configurations again. This implies that the lifetime of the protoplanetary disc needs to be long enough for this to happen. In contrast to the formation channel in viscously heated disc, instabilities are here required to reach the observed mass distribution. These instabilities can cause high mutual inclinations between the planets \citep{2017MNRAS.470.1750I, 2019arXiv190208772I}. This implies that combined observations of transits and RV detections can disentangle the two formation channels by determining how many planets are hidden between observed transiting planets, giving important constraints on instabilities of super-Earths systems in the inner disc. Additionally this could imply that the final structure of super-Earth systems reveals information if viscous heating in the inner protoplanetary disc is important.

\section{Discussion}
\label{sec:discussion}

In this section I list the four main assumptions in the work. I also discuss their implication on the formation of rocky super-Earths within the first Myr of disc evolution.

\subsection{Protoplanetary disc evolution}

The structure of the protoplanetary disc sets the pebble isolation mass and has thus great influence on the formation of super-Earths. The here used viscous disc model \citep{2015A&A...575A..28B} follows the decrease in the disc's accretion rate over time following observations \citep{1998ApJ...495..385H}. This decrease in the disc's accretion rate over time can be explained by viscous evolution \citep{2018MNRAS.474...88H}. In addition, the viscous evolution models of \citet{2015arXiv150303352B} find a very similar reduction in the disc's accretion rate and thermal structure over time. Thus discs that generate enough viscous heating should follow similar structures and evolutions.

However, \citet{2019ApJ...872...98M} showed with MHD shearing box simulations that viscous heating in the disc midplane should be suppressed due to the low level of turbulence. This implies that the discs are quite cold and should only feature low pebble isolation masses at all time, in contradiction to the here used disc evolution models with viscous heating. However, a very cold disc at all times implies that the planetesimals and planetary embryos that formed the Earth in our own solar system must have been water rich, in contrast to the current theories of the Earth's formation. This implies that at least within the first Myr the inner disc of the solar system should have been hotter (e.g. by infall heating, current sheats, see \citealt{2014ApJ...791...62M}) and should thus have featured a larger pebble isolation mass. 

In the N-body simulations of \citet{2019arXiv190208772I}, the disc's lifetime was assumed to be 5 Myr, while typical protoplanetary discs seem to have lifetimes sorter than that \citep{2009AIPC.1158....3M}. A short disc lifetime reduces the time planets can migrate through gravitational interactions in the disc, which is important for the pile-up of super-Earths at the inner disc edge \citep{2017AJ....153..222B}.

\subsection{Planet migration and resonant systems}

Due the interactions with the gaseous protoplanetary disc, planets migrate through it \citep{1986Icar...67..164W}. Planetary migration for bodies of a few Earth masses happens on timescales shorter than the disc lifetime and is directed mostly inwards \citep{2002ApJ...565.1257T}. In regions with steep radial gradients of entropy, migration can be directed outwards \citep{2006A&A...459L..17P, 2008ApJ...672.1054B}, if the viscosity is high enough \citep{2002A&A...387..605M, 2013A&A...550A..52B}. This is the case in the here used viscously dominated disc model and planets of a few Earth masses can migrate outwards \citep{2015A&A...575A..28B}. Multiple planets piling up at close distances to each other increase their eccentricity through mutual interactions. However, this increase in eccentricity quenches the corotation torque and thus outward migration \citep{2010A&A.523...A30} and chains of planets migrate inwards \citep{2013A&A...553L...2C}. 

The migration of the forming planets is ultimately stopped at the inner edge of the disc, where the planets can pile up in chains of resonant planets \citep{2017MNRAS.470.1750I, 2019arXiv190208772I, 2019arXiv190208694L}. The migration speed of the planets sets in which resonance the planets are finally trapped, because migrating planets can skip weaker resonances easier if the migration speed is large. On the other hand, \citet{2019MNRAS.tmpL.120M} found that resonance trapping in discs with low viscosity might be much harder than in discs with large viscosity. However, they also find that the systems undergo a large instability rate after gas disc dissipation, consistent with the simulations of \citet{2017MNRAS.470.1750I, 2019arXiv190208772I, 2019arXiv190208694L}.

The planetary cores that form these chains of resonant systems during the gas disc phase have reached pebble isolation mass and consist mostly of bodies of a few Earth masses \citep{2019arXiv190208772I, 2019arXiv190208694L}. These chains can then after gas disc dissipation become unstable depending on the number of available embryos, their masses and their distances \citep{2006AJ....131.3093I, 2012Icar..221..624M, 2019A&A...625A...7P}. Planetary systems harbouring more massive and larger number of planets can become unstable on shorter timescales.

In total a fraction of 1-5$\%$ stable systems is needed to explain the observations of exoplanets \citep{2017MNRAS.470.1750I, 2019arXiv190208772I}. As shown in Fig.~\ref{fig:Masses} the mass distribution of the systems that are close to mean motion resonances is very similar to the distribution of systems that are outside of mean motion resonances. This leads to the describtion of the two formation channels for close-in super-Earths (section~\ref{sec:chanel}) for planets forming in viscous or passively heated discs.

In the systems formed in passively heated discs or old viscously heated discs, collisions can increase the masses of the planets, where this increase is about a factor of 2-3 from their original mass \citep{2019arXiv190208772I}. Planetary systems that form in these discs need to have multiple mergers to grow to masses large enough to explain the Kepler observations. In \citet{2019arXiv190208772I} the planets forming at late times reach low pebble isolation masses and can through mergers produce the Kepler observations regarding planetary masses. However, they do not reproduce a sufficient amount of instabilities to explain the Kepler observations. On the other hand, these simulations still lack mechanisms like tides or leftover planetesimals \citep{2005Natur.435..459T} that could trigger instabilities. 

In contrast, planets formed in viscously heated discs within 1 Myr should not undergo multiple mergers as their planetary masses would become too large for the Kepler observations \citep{2019arXiv190208772I}. However it is yet unclear which mechanism could drive this breaking of the resonant chains without causing major instabilities among the planets (e.g. the magnetic rebound effect \citep{2017A&A...606A..66L, 2017A&A...601A..15L} has not been tested for systems with many planets in a resonant chains). On the other hand, young stars rotate rapidly, generating a significant quadrupole moment, which torques the planetary orbits, with inner planets influenced more strongly, explaining the apparent large number of observations of single systems without invoking large instabilities \citep{2016ApJ...830....5S}.

\subsection{Gas accretion}

Planetary cores of a few Earth masses could in principle start to undergo gas accretion, which would allow them to increase their mass well above the pebble isolation mass,  which is why the maximum mass in Fig.~\ref{fig:Masses} is set to 10 Earth masses as planets close to this mass could start runaway gas accretion \citep{1996Icar..124...62P}. However, the gas accretion process is hindered by recycling flows of the gas that penetrate into the Hill sphere of the planets, making gas accretion onto planets of just a few Earth masses very difficult \citep{2015MNRAS.446.1026O, 2015MNRAS.447.3512O, 2017A&A...606A.146L, 2017MNRAS.471.4662C}. In addition, the envelope contraction depends on the opacity within the envelope \citep{2014A&A...572A.118M, 2014ApJ...786...21P}. It was suggested by \citet{2017A&A...606A.146L} that the opacity in the inner disc might be larger than in the outer disc, implying that planets in the inner regions do not grow to gas giants. However, the exact opacities in the planetary envelope and thus the contraction and gas accretion rates depend on the underlying pebble sizes, compositions and infall rates into the envelope. Additionally, leftover planetesimals could be ablated in the planetary envelope, heating it and thus preventing contraction \citep{2018NatAs...2..873A}.

Planets close to their central star can lose their atmosphere due to stellar photoevaporation \citep{2012MNRAS.425.2931O, 2013ApJ...775..105O, 2017ApJ...847...29O, 2018ApJ...853..163J} or by just evaporating their envelope due to cooling of the planetary core \citep{2019MNRAS.487...24G}. This mainly applies for planets up to a few Earth masses, indicating that the observed planetary radii could correspond to the bare rocky and/or icy planetary core. The final size and mass of these planets is thus determined only by their solid components and not by gas accretion. These arguments result in not modelling gas accretion onto small cores formed in the inner disc and to a cut at the maximum mass of 10 Earth masses in the here presented work.

\subsection{Planet composition}

Recent analysis of the Kepler data has revealed that close-in super-Earth could be predominantly of rocky composition \citep{2017ApJ...847...29O, 2018ApJ...853..163J}. A planetary embryo accreting pebbles interior to the water ice line ($r<r_{\rm ice}$) will by construction have no water ice content. Planets forming exterior to the water ice line might have instead a large water ice content \citep{2019arXiv190208772I, 2019A&A...624A.109B}, potentially inconsistent with the observations. This implies that planets forming in the outer disc (even at late times) and then migrating inwards after their formation is complete might have masses that match the observations, but their composition would be dominated by ice instead of rock\footnote{This applies if there is no other process that allows the planet to lose water ice during its formation, e.g. volatile loss of the pebbles during the accretion process.}.

However, some super-Earths might actually contain a significant water ice fraction \citep{2019arXiv190604253Z}, implying that these planets must have formed in the outer parts of the protoplanetary disc and then migrated inwards, as described in \citet{2019arXiv190208772I}. However, the study of \citet{2019arXiv190208772I} showed that planetary embryos growing exterior to the water ice line form systems with only water rich planets, due to the fast growth by icy pebbles in the outer disc. This thus implies that some fraction of the super-Earths must form in the inner regions of the protoplanetary disc. The upcoming RV follow up observations of the TESS mission will thus be of crucial importance to improve our understanding of the composition of super-Earths and thus their formation channels. I focus in this work on the formation of rocky super-Earths in the inner regions (up to 2-3 AU) of the protoplanetary disc.

The idea of \citet{2019A&A...624A.109B} to form the cores of super-Earths at the water ice line and then migrate them inwards where they finish to assemble in the inner disc results in low water ice content of these planets, presumably also consistent with observations and with the Trappist-1 system (see also \citealt{2019arXiv190600669S}). However, as the planets finish their formation in the inner disc, their growth is stopped at the pebble isolation mass in the inner disc, implying that planets still need to form early enough to reach high enough masses consistent with the Kepler observations (Fig.~\ref{fig:Masses}). This indicates that the formation of planetary embryos that then form the rocky super-Earths could happen interior or exterior to the water ice line, as long as their formation is not completed exterior to the water ice line.

\section{Summary}
\label{sec:summary}

By comparing the masses of observed planets with the pebble isolation mass in viscously heated discs and discs solely heated by their central star, I identified two formation channels for the formation of rocky close-in super-Earths via pebble accretion.

In the pebble accretion scenario, the growth of the planet is stopped at the pebble isolation mass (eq.~\ref{eq:MisowD}), which is when the planet opens a partial gap in the protoplanetary disc structure stopping the inward flow of pebbles. This pebble isolation mass is a strong function of the disc's thermal properties and thus changes as the disc evolves in time (Fig.~\ref{fig:Planets}). The pebble isolation mass is independent of the pebble flux, indicating that planets grow to similar masses.

After about 1 Myr of disc evolution, the pebble isolation mass in the inner disc is much smaller than the bulk of the observed planet population for viscously heated discs. This implies that the formation of super-Earths has to happen on shorter timescales to explain the masses of the observed exoplanet population if planetary mergers are not important for the final system architecture and mass. 

Due to type-I migration, planets migrate to the inner edge of the disc where they form systems of resonant planets, naturally explaining the systems in resonance. However, all systems formed in this way are then in resonant configurations. In order to break the resonant chains, instabilities should not play a major role, because this would increase the planetary masses to values larger than predicted by the Kepler observations. The resonant chains could then be broken through other mechanism, e.g. magnetic rebound \citep{2017A&A...606A..66L, 2017A&A...601A..15L} or the disc lifetime could be so short that the planets are not trapped in resonance in the first place \citep{2017AJ....153..222B} or they form in low viscosity environments where trapping of multiple planets in resonance seems much harder \citep{2019MNRAS.tmpL.120M}.

On the other hand, if the discs are not heated viscously, the pebble isolation mass in the inner regions of the disc is much smaller, implying that mutual mergers between the planets are needed to grow planets large enough to match the Kepler constraints. However, the resulting systems after a series of collisions would not be in a resonant configuration, if the collisions happen after the gas-disc phase \citep{2019arXiv190208772I, 2019arXiv190208694L}. In order to produce resonant systems by planetary mergers, these have to happen during the gas disc phase, so that the system can be trapped in resonance through type-I migration, as shown in \citet{2017MNRAS.470.1750I}. However, it is not trivial to produce systems purely by giant impacts that match the Kepler observations in respect to the similarity of planets within the system \citep{2018A&A...615A..63O}, but giant impacts could produce the size ratio distribution of adjacent planet pairs \citep{2018ApJ...866..104C}.

Independently if the inner disc structure is dominated by viscous heating or not, planetary growth via pebble accretion can easily happen within 1 Myr, mostly independent of the pebble flux (appendix~\ref{ap:growth}). The short formation time-scale also seems to avoid the general trend that the amount of available solids in protoplanetary discs decreases quite rapidly after 1 Myr, where on average only $\sim$10 Earth masses of pebbles are available \citep{2016ApJ...828...46A, 2017AJ....153..240A, 2019ApJ...875L...9W}, believed to not be enough to form planetary systems that have on average larger masses than that \citep{2018A&A...618L...3M}. However, the study by \citet{2019arXiv190402127Z} suggests that the dust masses could be higher by a factor even larger than 10 due to the optical thickness of the disc.

The here proposed two formation channels of close-in rocky super-Earths have different implications for the resulting planetary systems. If the planetary systems undergo massive instabilities to break their resonant configuration, the mutual inclination between the remaining planets is high enough to explain the overabundance of planetary systems where only one planet is detected by transits even though more planets could be hidden within the system \citep{2019arXiv190208772I}. In contrast, if the planetary chains do not undergo major instabilities and their resonant configuration is broken by other mechanisms that do not cause major instabilities, the number of planets hidden in transit observations could be small. The upcoming RV surveys to the transit detections around bright stars with TESS could help to distinguish between these two formation channels, because these surveys will improve greatly the statistics of hidden planets in planetary systems originally discovered by transits. Additionally, these observations will help to improve the mass-radius relationship for close-in super-Earths giving further constraints on the formation location of super-Earths.

\begin{acknowledgements}

B.B. thanks A. Izidoro, M. Lambrechts, and A. Johansen for fruitful discussions as well as an anonymous referee for her/his comments that helped to improve the manuscript. B.B. thanks the European Research Council (ERC Starting Grant 757448-PAMDORA) for their financial support.

\end{acknowledgements}

\appendix
\section{Pebble isolation mass}
\label{ap:Miso}

\begin{figure}
 \centering
 \includegraphics[scale=0.67]{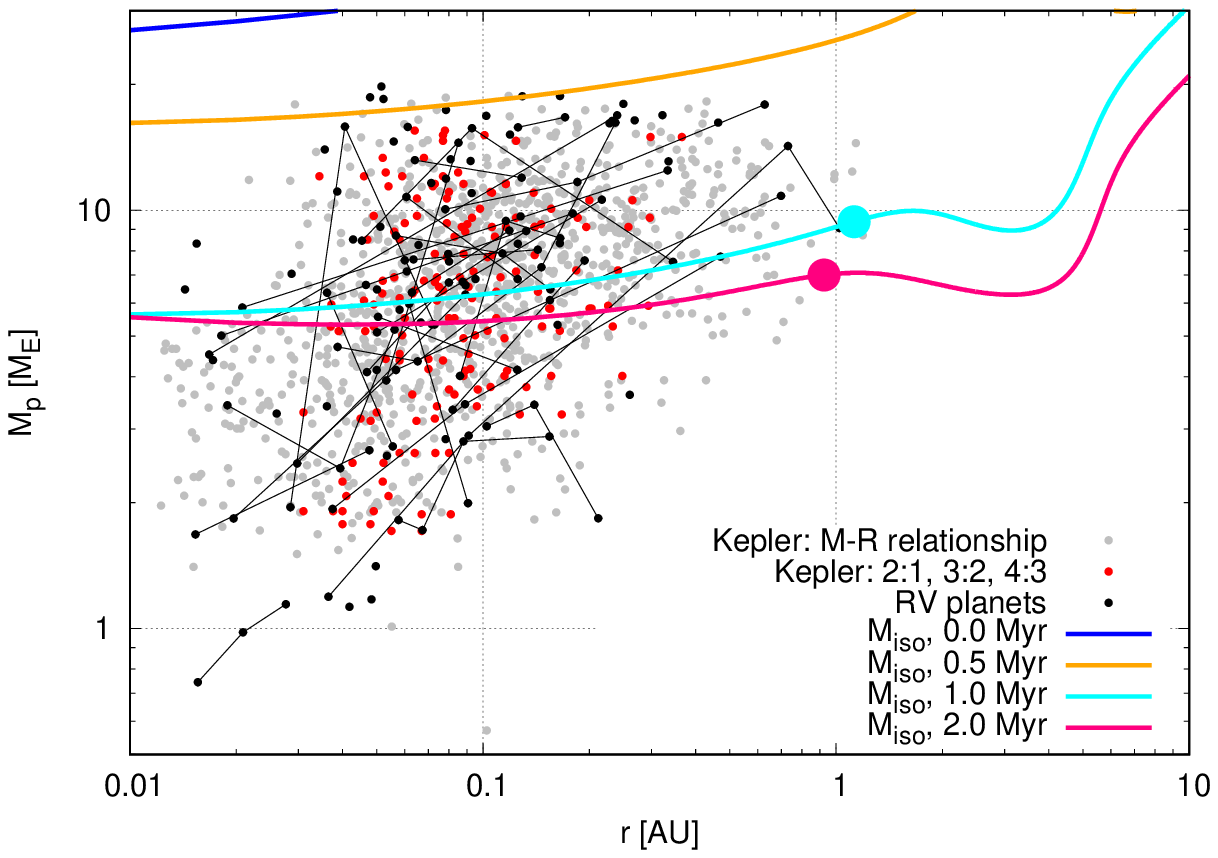}
 \includegraphics[scale=0.67]{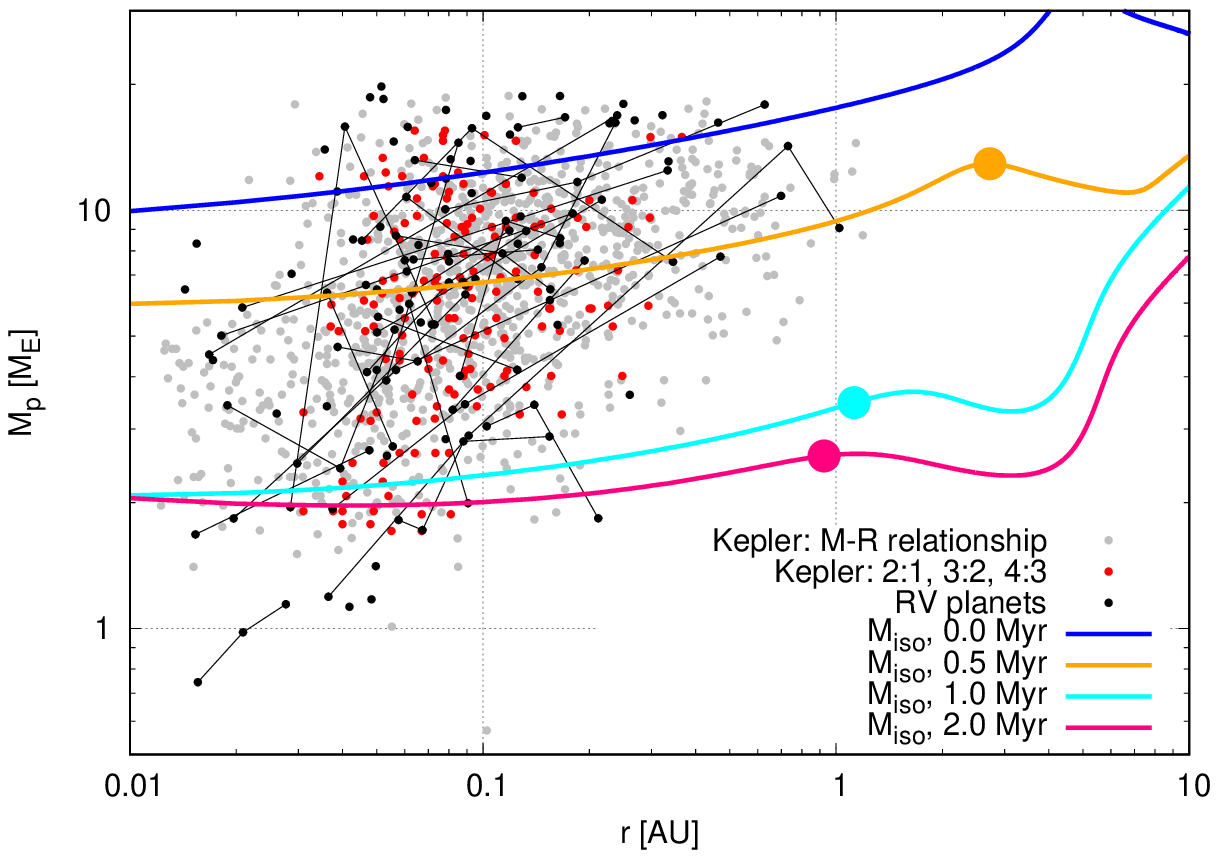} 
 \caption{Same as Fig.~\ref{fig:Planets}, but using $\alpha=0.0054$ (top) and $\alpha=10^{-4}$ (bottom) for the pebble isolation mass, but keeping the disc structure the same for the pebble isolation mass estimates. The change of the $\alpha$ parameter changes the pebble isolation mass, but the trend is similar for different $\alpha$ values: a large fraction of super-Earth planets are more massive than the pebble isolation mass when the disc reaches an age of 1 Myr.
   \label{fig:Miso}
   }
\end{figure}

Fig.~\ref{fig:Miso} the planetary masses from the mass radius-relationship of the Kepler observations and from RV observations (similar as in Fig.~\ref{fig:Planets}) are shown, but I used two different $\alpha$ values to calculate the pebble isolation mass. The overall disc structure remains unchanged for this calculation. The disc's thermal structure is calculated with $\alpha=0.0054$ \citet{2015A&A...575A..28B}. If the accretion rate of the disc is determined by $\dot{M}_{\rm disc} = 3 \pi \nu \Sigma_{\rm gas}$ an exchange of viscosity and surface density can give a similar thermal disc structure, even though the relationship is not 1:1. A disc with the same $\dot{M}_{\rm disc}$, but lower viscosity will have a larger temperature through viscous heating as a disc with higher viscosity, due to the disc's cooling which is proportional to $1/\rho_{\rm gas}$.

The pebble isolation mass scales only weakly with viscosity (eq.~\ref{eq:MisowD}), which changes the pebble isolation mass by a factor of two for high versus low viscosity. However, even for the high viscosity and thus high pebble isolation mass most super-Earths are larger than the pebble isolation mass for discs older than 1 Myr. In addition, detailed simulation of the disc's turbulence show low level of turbulence \citep{2013ApJ...769...76B, 2013ApJ...772...96B, 2016arXiv160900437S}, which is also found in observations \citep{2018ApJ...856..117F, 2018ApJ...869L..46D}. This implies that the trend described in the main paper remains valid and is mainly independent on the disc's viscosity.

\section{Pebble growth and planet migration}
\label{ap:growth}

\begin{figure}
 \centering
 \includegraphics[scale=0.7]{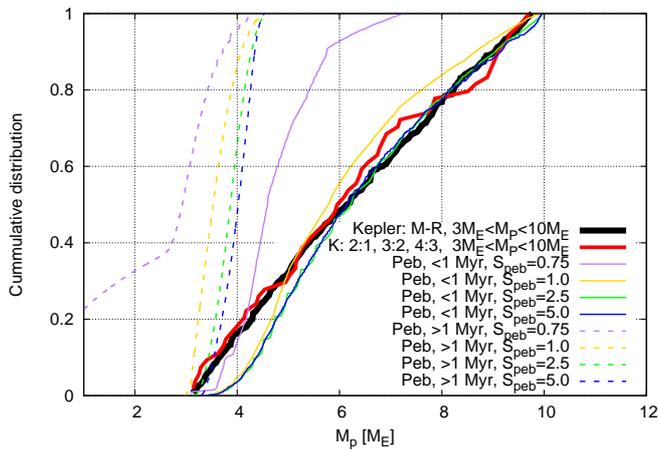} 
 \caption{Same as fig.~\ref{fig:Masses}, but for different pebble flux scalings $S_{\rm peb}$ and only for the viscously heated disc. Only for very low pebble fluxes do planets not reach the pebble isolation mass any more and they thus have to additional grow through collisions to match the masses of super-Earths determined by the Kepler mission.
   \label{fig:Masses2}
   }
\end{figure}

I use here the same pebble growth mechanism as outlined in \citet{2019arXiv190208772I} and \citet{2019A&A...623A..88B} to make the comparison to the super-Earths formation simulations of \citet{2019arXiv190208772I} easier. Thus only the necessities are repeated and I show in Fig.~\ref{fig:Masses2} the cumulative distribution of planetary masses for super-Earths grown in viscously heated discs where planets grow with different pebble fluxes.

Pebbles in the protoplanetary disc settle towards the midplane depending on their size, parameterised in this work by the dimensionless Stokes number $\tau_{\rm f}$, and depending on the level of turbulence in the protoplanetary disc described through the disc's viscosity. Using the $\alpha$ prescription, \citet{2007Icar..192..588Y} derived the pebble scale height as
\begin{equation}
 H_{\rm peb} = H_{\rm g} \sqrt{\alpha_{\rm disc} / \tau_{\rm f}} \ .
\end{equation}
Typically the pebbles in the here presented simulations have Stokes numbers of 0.05-0.2, which is calculated by equating the drift time-scale with the growth time-scale \citep{2012A&A...539A.148B, 2012A&A...544A..32L}. That yields a value of $H_{\rm peb}/H_{\rm g}\sim$0.1, in agreement with observations of protoplanetary discs \citep{2016ApJ...816...25P}.

The accretion rate onto the planet is determined by the pebble surface density $\Sigma_{\rm peb}$ and the planetary Hill radius $r_{\rm H}$ through the following equation
\begin{equation}
 \dot{M}_{\rm core} = 2 \left( \frac{\tau_f}{0.1} \right)^{2/3} r_{\rm H}^2 \Omega_{\rm K} \Sigma_{\rm peb} (r_{\rm P}) \ ,
\end{equation}
where $\Omega_{\rm K}$ is the Keplerian frequency. The pebble surface density $\Sigma_{\rm peb} (r_{\rm P})$ at the planets location can be calculated from the pebble flux $\dot{M}_{\rm peb}$ via
\begin{equation}
 \label{eq:SigmaPeb}
 \Sigma_{\rm peb} (r_{\rm P}) = \sqrt{\frac{2 S_{\rm peb} \dot{M}_{\rm peb} \Sigma_{\rm g}(r_{\rm P}) }{\sqrt{3} \pi \epsilon_{\rm P} r_{\rm P} v_{\rm K}}} \ ,
\end{equation}
where $r_{\rm P}$ denotes the semi-major axis of the planet, $v_{\rm K} = \Omega_{\rm K} r_{\rm P}$, and $\Sigma_{\rm g} (r_{\rm P})$ stands for the gas surface density at the planets locations. The pebble flux $\dot{M}_{\rm peb}$ is calculated self consistently through an equilibrium between dust growth and drift \citep{2012A&A...539A.148B, 2014A&A...572A.107L, 2018A&A...609C...2B}, where these simulations predict a decrease of the pebble flux in time. Here $S_{\rm peb}$ describes the scaling factor to the pebble flux $\dot{M}_{\rm peb}$ to test the influence of different pebble fluxes (see below). The pebble sticking efficiency can be taken as $\epsilon_{\rm P} =0.5$ under the assumption of near-perfect sticking \citep{2014A&A...572A.107L}. 

The Stokes number of the pebbles can be related to the pebble surface density $\Sigma_{\rm peb}$ and gas surface density $\Sigma_{\rm g}$ through the following relation
\begin{equation}
 \tau_{\rm f} = \frac{\sqrt{3}}{8} \frac{\epsilon_{\rm P}}{\eta} \frac{\Sigma_{\rm peb}(r_{\rm P})}{\Sigma_{\rm g} (r_{\rm P})} \ .
\end{equation}
Here $\eta$ represents a measurement of the sub-Keplerianity of the gas velocity.

The effects of an outer reservoir of pebbles in the disc as proposed in \citet{2018A&A...609C...2B} are not included, but I follow the reduction of the pebble flux in time as predicted directly by the pebble evolution models. This yields $\dot{M}_{\rm peb}$ and thus $\Sigma_{\rm peb}$ of values lower than predicted by observations \citep{2018A&A...609C...2B}. These low $\Sigma_{\rm peb}$ values result in very low pebble accretion rates, making the formation of giant planet cores difficult. I use thus a scaling factor $S_{\rm peb}$ to increase the pebble flux. Independently of the exact pebble accretion model used, the trend is clear, namely that only high enough pebble fluxes allow the growth to pebble isolation mass within 1 Myr of disc evolution.

The migration of the planet is modelled following the type-I prescription of \citet{2011MNRAS.410..293P}. In this prescription, the planets can migrate outwards, if the radial gradients of entropy are steep. This normally happens exterior to the water ice line \citep{2013A&A...549A.124B, 2014A&A...564A.135B, 2015A&A...575A..28B}. In this case, the planets are formed interior to the water ice line and thus migrate mostly inwards. If the planets reach the inner edge of the disc at 0.04 AU, the planet is trapped there and it is allowed to continue to accrete until it reaches the pebble isolation mass.

\bibliographystyle{aa}
\bibliography{Stellar}

\begin{thebibliography}{95}
\expandafter\ifx\csname natexlab\endcsname\relax\def\natexlab#1{#1}\fi

\bibitem[{{Alibert} {et~al.}(2018){Alibert}, {Venturini}, {Helled}, {Ataiee},
  {Burn}, {Senecal}, {Benz}, {Mayer}, {Mordasini}, \&
  {Quanz}}]{2018NatAs...2..873A}
{Alibert}, Y., {Venturini}, J., {Helled}, R., {et~al.} 2018, Nature Astronomy,
  2, 873

\bibitem[{Ansdell {et~al.}(2017)Ansdell, Williams, Manara, Miotello, Facchini,
  van~der Marel, Testi, \& van Dishoeck}]{2017AJ....153..240A}
Ansdell, M., Williams, J.~P., Manara, C.~F., {et~al.} 2017, AJ, 153, id. 240

\bibitem[{{Ansdell} {et~al.}(2016){Ansdell}, {Williams}, {van der Marel},
  {Carpenter}, {Guidi}, {Hogerheijde}, {Mathews}, {Manara}, {Miotello},
  {Natta}, {Oliveira}, {Tazzari}, {Testi}, {van Dishoeck}, \& {van
  Terwisga}}]{2016ApJ...828...46A}
{Ansdell}, M., {Williams}, J.~P., {van der Marel}, N., {et~al.} 2016, \apj,
  828, 46

\bibitem[{{Ataiee} {et~al.}(2018){Ataiee}, {Baruteau}, {Alibert}, \&
  {Benz}}]{2018A&A...615A.110A}
{Ataiee}, S., {Baruteau}, C., {Alibert}, Y., \& {Benz}, W. 2018, \aap, 615,
  A110

\bibitem[{Bai(2013)}]{2013ApJ...772...96B}
Bai, X.~N. 2013, ApJ, 772, id.96

\bibitem[{Bai \& Stone(2013)}]{2013ApJ...769...76B}
Bai, X.~N. \& Stone, J.~M. 2013, ApJ, 769, id. 76

\bibitem[{Bailli{\'e} {et~al.}(2015)Bailli{\'e}, Charnoz, \&
  Pantin}]{2015arXiv150303352B}
Bailli{\'e}, K., Charnoz, S., \& Pantin, {\'E}. 2015, astro-ph.EP

\bibitem[{{Baraffe} {et~al.}(1998){Baraffe}, {Chabrier}, {Allard}, \&
  {Hauschildt}}]{1998A&A...337..403B}
{Baraffe}, I., {Chabrier}, G., {Allard}, F., \& {Hauschildt}, P.~H. 1998, A\&A,
  337, p.403

\bibitem[{{Baraffe} {et~al.}(2015){Baraffe}, {Homeier}, {Allard}, \&
  {Chabrier}}]{2015A&A...577A..42B}
{Baraffe}, I., {Homeier}, D., {Allard}, F., \& {Chabrier}, G. 2015, \aap, 577,
  A42

\bibitem[{{Baruteau} \& {Masset}(2008)}]{2008ApJ...672.1054B}
{Baruteau}, C. \& {Masset}, F. 2008, \apj, 672, 1054

\bibitem[{Birnstiel {et~al.}(2012)Birnstiel, Klahr, \&
  Ercolano}]{2012A&A...539A.148B}
Birnstiel, T., Klahr, H., \& Ercolano, B. 2012, A\&A, 539, id.A148

\bibitem[{{Bitsch} {et~al.}(2013{\natexlab{a}}){Bitsch}, {Boley}, \&
  {Kley}}]{2013A&A...550A..52B}
{Bitsch}, B., {Boley}, A.~C., \& {Kley}, W. 2013{\natexlab{a}}, A\&A, 550,
  id.A52

\bibitem[{{Bitsch} {et~al.}(2013{\natexlab{b}}){Bitsch}, {Crida}, {Morbidelli},
  {Kley}, \& {Dobbs-Dixon}}]{2013A&A...549A.124B}
{Bitsch}, B., {Crida}, A., {Morbidelli}, A., {Kley}, W., \& {Dobbs-Dixon}, I.
  2013{\natexlab{b}}, A\&A, 549, id.A124

\bibitem[{{Bitsch} {et~al.}(2019{\natexlab{a}}){Bitsch}, {Izidoro}, {Johansen},
  {Raymond}, {Morbidelli}, {Lambrechts}, \& {Jacobson}}]{2019A&A...623A..88B}
{Bitsch}, B., {Izidoro}, A., {Johansen}, A., {et~al.} 2019{\natexlab{a}}, \aap,
  623, A88

\bibitem[{Bitsch {et~al.}(2015{\natexlab{a}})Bitsch, Johansen, Lambrechts, \&
  Morbidelli}]{2015A&A...575A..28B}
Bitsch, B., Johansen, A., Lambrechts, M., \& Morbidelli, A. 2015{\natexlab{a}},
  A\&A, 575, id.A28

\bibitem[{{Bitsch} \& {Kley}(2010)}]{2010A&A.523...A30}
{Bitsch}, B. \& {Kley}, W. 2010, \aap, 523, A30

\bibitem[{Bitsch {et~al.}(2015{\natexlab{b}})Bitsch, Lambrechts, \&
  Johansen}]{2015A&A...582A.112B}
Bitsch, B., Lambrechts, M., \& Johansen, A. 2015{\natexlab{b}}, A\&A, 582,
  id.A112

\bibitem[{{Bitsch} {et~al.}(2018{\natexlab{a}}){Bitsch}, {Lambrechts}, \&
  {Johansen}}]{2018A&A...609C...2B}
{Bitsch}, B., {Lambrechts}, M., \& {Johansen}, A. 2018{\natexlab{a}}, \aap,
  609, C2

\bibitem[{{Bitsch} {et~al.}(2018{\natexlab{b}}){Bitsch}, {Morbidelli},
  {Johansen}, {Lega}, {Lambrechts}, \& {Crida}}]{2018arXiv180102341B}
{Bitsch}, B., {Morbidelli}, A., {Johansen}, A., {et~al.} 2018{\natexlab{b}},
  A\&A, 612, id.A30

\bibitem[{{Bitsch} {et~al.}(2014){Bitsch}, {Morbidelli}, {Lega}, \&
  {Crida}}]{2014A&A...564A.135B}
{Bitsch}, B., {Morbidelli}, A., {Lega}, E., \& {Crida}, A. 2014, A\&A, 564,
  id.A135

\bibitem[{{Bitsch} {et~al.}(2019{\natexlab{b}}){Bitsch}, {Raymond}, \&
  {Izidoro}}]{2019A&A...624A.109B}
{Bitsch}, B., {Raymond}, S.~N., \& {Izidoro}, A. 2019{\natexlab{b}}, \aap, 624,
  A109

\bibitem[{Brasser {et~al.}(2017)Brasser, Bitsch, \&
  Matsumura}]{2017AJ....153..222B}
Brasser, R., Bitsch, B., \& Matsumura, S. 2017, AJ, 153, id. 222

\bibitem[{Brauer {et~al.}(2008)Brauer, Dullemond, \&
  Henning}]{2008A&A...480..859B}
Brauer, F., Dullemond, C., \& Henning, T. 2008, A\&A, 480, pp.859

\bibitem[{{Carrera} {et~al.}(2018){Carrera}, {Ford}, {Izidoro},
  {Jontof-Hutter}, {Raymond}, \& {Wolfgang}}]{2018ApJ...866..104C}
{Carrera}, D., {Ford}, E.~B., {Izidoro}, A., {et~al.} 2018, \apj, 866, 104

\bibitem[{{Chiang} \& {Laughlin}(2013)}]{2013MNRAS.431.3444C}
{Chiang}, E. \& {Laughlin}, G. 2013, \mnras, 431, 3444

\bibitem[{{Chiang} \& {Youdin}(2010)}]{2010AREPS..38..493C}
{Chiang}, E. \& {Youdin}, A. 2010, Annual Review of Earth and Planetary
  Sciences, 38, p.493

\bibitem[{{Chiang} \& {Goldreich}(1997)}]{1997ApJ...490..368C}
{Chiang}, E.~I. \& {Goldreich}, P. 1997, ApJ, 490, 368

\bibitem[{{Cimerman} {et~al.}(2017){Cimerman}, {Kuiper}, \&
  {Ormel}}]{2017MNRAS.471.4662C}
{Cimerman}, N.~P., {Kuiper}, R., \& {Ormel}, C.~W. 2017, \mnras, 471, 4662

\bibitem[{{Cossou} {et~al.}(2013){Cossou}, {Raymond}, \&
  {Pierens}}]{2013A&A...553L...2C}
{Cossou}, C., {Raymond}, S.~N., \& {Pierens}, A. 2013, A\&A, 553, id.L2

\bibitem[{{Dullemond} {et~al.}(2018){Dullemond}, {Birnstiel}, {Huang},
  {Kurtovic}, {Andrews}, {Guzm{\'a}n}, {P{\'e}rez}, {Isella}, {Zhu}, {Benisty},
  {Wilner}, {Bai}, {Carpenter}, {Zhang}, \& {Ricci}}]{2018ApJ...869L..46D}
{Dullemond}, C.~P., {Birnstiel}, T., {Huang}, J., {et~al.} 2018, \apjl, 869,
  L46

\bibitem[{{Dullemond} \& Dominik(2004)}]{2004A&A...417..159D}
{Dullemond}, C.~P. \& Dominik, C. 2004, A\&A, 417, 159

\bibitem[{{Flaherty} {et~al.}(2018){Flaherty}, {Hughes}, {Teague}, {Simon},
  {Andrews}, \& {Wilner}}]{2018ApJ...856..117F}
{Flaherty}, K.~M., {Hughes}, A.~M., {Teague}, R., {et~al.} 2018, \apj, 856, 117

\bibitem[{{Gupta} \& {Schlichting}(2019)}]{2019MNRAS.487...24G}
{Gupta}, A. \& {Schlichting}, H.~E. 2019, \mnras, 487, 24

\bibitem[{{Hartmann} \& {Bae}(2018)}]{2018MNRAS.474...88H}
{Hartmann}, L. \& {Bae}, J. 2018, \mnras, 474, 88

\bibitem[{{Hartmann} {et~al.}(1998){Hartmann}, {Calvet}, {Gullbring}, \&
  {D'Alessio}}]{1998ApJ...495..385H}
{Hartmann}, L., {Calvet}, N., {Gullbring}, E., \& {D'Alessio}, P. 1998, ApJ,
  495, p.385

\bibitem[{{Ida} {et~al.}(2016){Ida}, {Guillot}, \&
  {Morbidelli}}]{2016A&A...591A..72I}
{Ida}, S., {Guillot}, T., \& {Morbidelli}, A. 2016, \aap, 591, A72

\bibitem[{{Iwasaki} \& {Ohtsuki}(2006)}]{2006AJ....131.3093I}
{Iwasaki}, K. \& {Ohtsuki}, K. 2006, \aj, 131, 3093

\bibitem[{{Izidoro} {et~al.}(2019){Izidoro}, {Bitsch}, {Raymond}, {Johansen},
  {Morbidelli}, {Lambrechts}, \& {Jacobson}}]{2019arXiv190208772I}
{Izidoro}, A., {Bitsch}, B., {Raymond}, S.~N., {et~al.} 2019, arXiv e-prints

\bibitem[{Izidoro {et~al.}(2017)Izidoro, Ogihara, Raymond, Morbidelli, Pierens,
  Bitsch, Cossou, \& Hersant}]{2017MNRAS.470.1750I}
Izidoro, A., Ogihara, M., Raymond, S.~N., {et~al.} 2017, MNRAS, 470, pp. 1750

\bibitem[{{Jin} \& {Mordasini}(2018)}]{2018ApJ...853..163J}
{Jin}, S. \& {Mordasini}, C. 2018, \apj, 853, 163

\bibitem[{Johansen \& Lacerda(2010)}]{2010MNRAS.404..475J}
Johansen, A. \& Lacerda, P. 2010, MNRAS, 404, pp. 475

\bibitem[{Kruijer {et~al.}(2017)Kruijer, Kleine, Burkhardt, \&
  Budde}]{2017LPI....48.1386K}
Kruijer, T.~S., Kleine, T., Burkhardt, C., \& Budde, G. 2017, LPI, 48th
  Conference, id.1386

\bibitem[{{Lambrechts} \& {Johansen}(2012)}]{2012A&A...544A..32L}
{Lambrechts}, M. \& {Johansen}, A. 2012, A\&A, 544, id.A32

\bibitem[{Lambrechts \& Johansen(2014)}]{2014A&A...572A.107L}
Lambrechts, M. \& Johansen, A. 2014, A\&A, 572, id.A107

\bibitem[{Lambrechts {et~al.}(2014)Lambrechts, Johansen, \&
  Morbidelli}]{2014A&A...572A..35L}
Lambrechts, M., Johansen, A., \& Morbidelli, A. 2014, A\&A, 572, id. A35

\bibitem[{{Lambrechts} \& {Lega}(2017)}]{2017A&A...606A.146L}
{Lambrechts}, M. \& {Lega}, E. 2017, \aap, 606, A146

\bibitem[{{Lambrechts} {et~al.}(2019){Lambrechts}, {Morbidelli}, {Jacobson},
  {Johansen}, {Bitsch}, {Izidoro}, \& {Raymond}}]{2019arXiv190208694L}
{Lambrechts}, M., {Morbidelli}, A., {Jacobson}, S.~A., {et~al.} 2019, arXiv
  e-prints

\bibitem[{{Liu} \& {Ormel}(2017)}]{2017A&A...606A..66L}
{Liu}, B. \& {Ormel}, C.~W. 2017, \aap, 606, A66

\bibitem[{{Liu} {et~al.}(2017){Liu}, {Ormel}, \& {Lin}}]{2017A&A...601A..15L}
{Liu}, B., {Ormel}, C.~W., \& {Lin}, D. N.~C. 2017, \aap, 601, A15

\bibitem[{{Lynden-Bell} \& {Pringle}(1974)}]{1974MNRAS.168..603L}
{Lynden-Bell}, D. \& {Pringle}, J.~E. 1974, \mnras, 168, 603

\bibitem[{Mamajek(2009)}]{2009AIPC.1158....3M}
Mamajek, E.~E. 2009, AIP Conference Proceedings, 1158, pp.3

\bibitem[{{Manara} {et~al.}(2018){Manara}, {Morbidelli}, \&
  {Guillot}}]{2018A&A...618L...3M}
{Manara}, C.~F., {Morbidelli}, A., \& {Guillot}, T. 2018, \aap, 618, L3

\bibitem[{{Masset}(2002)}]{2002A&A...387..605M}
{Masset}, F.~S. 2002, \aap, 387, 605

\bibitem[{{Matsumoto} {et~al.}(2012){Matsumoto}, {Nagasawa}, \&
  {Ida}}]{2012Icar..221..624M}
{Matsumoto}, Y., {Nagasawa}, M., \& {Ida}, S. 2012, \icarus, 221, 624

\bibitem[{{Mayor} {et~al.}(2011){Mayor}, {Marmier}, {Lovis}, {Udry},
  {S{\'e}gransan}, {Pepe}, {Benz}, {Bertaux}, {Bouchy}, {Dumusque}, {Lo Curto},
  {Mordasini}, {Queloz}, \& {Santos}}]{2011arXiv1109.2497M}
{Mayor}, M., {Marmier}, M., {Lovis}, C., {et~al.} 2011, ArXiv e-prints

\bibitem[{McNally {et~al.}(2014)McNally, Hubbard, Yang, \& {Mac
  Low}}]{2014ApJ...791...62M}
McNally, C.~P., Hubbard, A., Yang, C.-C., \& {Mac Low}, M.~M. 2014, ApJ, 791,
  id. 62

\bibitem[{{McNally} {et~al.}(2019){McNally}, {Nelson}, \&
  {Paardekooper}}]{2019MNRAS.tmpL.120M}
{McNally}, C.~P., {Nelson}, R.~P., \& {Paardekooper}, S.-J. 2019, \mnras, L120

\bibitem[{{Morbidelli} \& {Nesvorny}(2012)}]{2012A&A...546A..18M}
{Morbidelli}, A. \& {Nesvorny}, D. 2012, A\&A, 546, id.A18

\bibitem[{Mordasini(2014)}]{2014A&A...572A.118M}
Mordasini, C. 2014, A\&A, 572, id.A118

\bibitem[{{Mori} {et~al.}(2019){Mori}, {Bai}, \&
  {Okuzumi}}]{2019ApJ...872...98M}
{Mori}, S., {Bai}, X.-N., \& {Okuzumi}, S. 2019, \apj, 872, 98

\bibitem[{{Mulders} {et~al.}(2018){Mulders}, {Pascucci}, {Apai}, \&
  {Ciesla}}]{2018AJ....156...24M}
{Mulders}, G.~D., {Pascucci}, I., {Apai}, D., \& {Ciesla}, F.~J. 2018, \aj,
  156, 24

\bibitem[{{Ndugu} {et~al.}(2019){Ndugu}, {Bitsch}, \&
  {Jurua}}]{2019MNRAS.tmp.1807N}
{Ndugu}, N., {Bitsch}, B., \& {Jurua}, E. 2019, \mnras, 1807

\bibitem[{{Ogihara} {et~al.}(2018){Ogihara}, {Kokubo}, {Suzuki}, \&
  {Morbidelli}}]{2018A&A...615A..63O}
{Ogihara}, M., {Kokubo}, E., {Suzuki}, T.~K., \& {Morbidelli}, A. 2018, \aap,
  615, A63

\bibitem[{{Ogihara} {et~al.}(2015){Ogihara}, {Morbidelli}, \&
  {Guillot}}]{2015A&A...578A..36O}
{Ogihara}, M., {Morbidelli}, A., \& {Guillot}, T. 2015, \aap, 578, A36

\bibitem[{{Oka} {et~al.}(2011){Oka}, {Nakamoto}, \&
  {Ida}}]{2011ApJ...738..141O}
{Oka}, A., {Nakamoto}, T., \& {Ida}, S. 2011, ApJ, 738, id.141

\bibitem[{Ormel \& Klahr(2010)}]{2010A&A...520A..43O}
Ormel, C.~W. \& Klahr, H.~H. 2010, A\&A, 520, id.A43

\bibitem[{{Ormel} {et~al.}(2015){Ormel}, {Kuiper}, \&
  {Shi}}]{2015MNRAS.446.1026O}
{Ormel}, C.~W., {Kuiper}, R., \& {Shi}, J.-M. 2015, \mnras, 446, 1026

\bibitem[{Ormel {et~al.}(2015)Ormel, Shi, \& Kuiper}]{2015MNRAS.447.3512O}
Ormel, C.~W., Shi, J.-M., \& Kuiper, R. 2015, MNRAS, 447, p.3512

\bibitem[{Owen \& Jackson(2012)}]{2012MNRAS.425.2931O}
Owen, J.~E. \& Jackson, A.~P. 2012, MNRAS, 425, pp.2931

\bibitem[{{Owen} \& {Wu}(2013)}]{2013ApJ...775..105O}
{Owen}, J.~E. \& {Wu}, Y. 2013, \apj, 775, 105

\bibitem[{{Owen} \& {Wu}(2017)}]{2017ApJ...847...29O}
{Owen}, J.~E. \& {Wu}, Y. 2017, \apj, 847, 29

\bibitem[{{Paardekooper} {et~al.}(2011){Paardekooper}, {Baruteau}, \&
  {Kley}}]{2011MNRAS.410..293P}
{Paardekooper}, S.~J., {Baruteau}, C., \& {Kley}, W. 2011, MNRAS, 410, 293

\bibitem[{Paardekooper \& Mellema(2006)}]{2006A&A...453.1129P}
Paardekooper, S.-J. \& Mellema, G. 2006, A\&A, 453, pp.1129

\bibitem[{{Paardekooper} \& {Mellema}(2006)}]{2006A&A...459L..17P}
{Paardekooper}, S.-J. \& {Mellema}, G. 2006, \aap, 459, L17

\bibitem[{{Pichierri} {et~al.}(2019){Pichierri}, {Batygin}, \&
  {Morbidelli}}]{2019A&A...625A...7P}
{Pichierri}, G., {Batygin}, K., \& {Morbidelli}, A.~r. 2019, \aap, 625, A7

\bibitem[{{Pinte} {et~al.}(2016){Pinte}, {Dent}, {M{\'e}nard}, {Hales}, {Hill},
  {Cortes}, \& {de Gregorio-Monsalvo}}]{2016ApJ...816...25P}
{Pinte}, C., {Dent}, W.~R.~F., {M{\'e}nard}, F., {et~al.} 2016, \apj, 816, 25

\bibitem[{Piso \& Youdin(2014)}]{2014ApJ...786...21P}
Piso, A.~M.~A. \& Youdin, A. 2014, ApJ, 786, id. 21

\bibitem[{{Pollack} {et~al.}(1996){Pollack}, {Hubickyj}, {Bodenheimer},
  {Lissauer}, {Podolak}, \& {Greenzweig}}]{1996Icar..124...62P}
{Pollack}, J.~B., {Hubickyj}, O., {Bodenheimer}, P., {et~al.} 1996, Icarus,
  124, 62

\bibitem[{{Raymond} \& {Cossou}(2014)}]{2014MNRAS.440L..11R}
{Raymond}, S.~N. \& {Cossou}, C. 2014, \mnras, 440, L11

\bibitem[{{Schoonenberg} {et~al.}(2019){Schoonenberg}, {Liu}, {Ormel}, \&
  {Dorn}}]{2019arXiv190600669S}
{Schoonenberg}, D., {Liu}, B., {Ormel}, C.~W., \& {Dorn}, C. 2019, arXiv
  e-prints, arXiv:1906.00669

\bibitem[{{Spalding} \& {Batygin}(2016)}]{2016ApJ...830....5S}
{Spalding}, C. \& {Batygin}, K. 2016, \apj, 830, 5

\bibitem[{Suzuki {et~al.}(2016)Suzuki, Ogihara, Morbidelli, Crida, \&
  Guillot}]{2016arXiv160900437S}
Suzuki, T.~K., Ogihara, M., Morbidelli, A., Crida, A., \& Guillot, T. 2016,
  A\&A, 596, id.A74

\bibitem[{{Tanaka} {et~al.}(2002){Tanaka}, {Takeuchi}, \&
  {Ward}}]{2002ApJ...565.1257T}
{Tanaka}, H., {Takeuchi}, T., \& {Ward}, W.~R. 2002, \apj, 565, 1257

\bibitem[{Tsiganis {et~al.}(2005)Tsiganis, Gomes, Morbidelli, \&
  Levison}]{2005Natur.435..459T}
Tsiganis, K., Gomes, R., Morbidelli, A., \& Levison, H.~F. 2005, Nature, 435,
  pp.459

\bibitem[{{Ward}(1986)}]{1986Icar...67..164W}
{Ward}, W.~R. 1986, Icarus, 67, 164

\bibitem[{Weidenschilling(1977)}]{1977MNRAS.180...57W}
Weidenschilling, S.~J. 1977, MNRAS, 180, p.57

\bibitem[{{Weiss} {et~al.}(2018){Weiss}, {Marcy}, {Petigura}, {Fulton},
  {Howard}, {Winn}, {Isaacson}, {Morton}, {Hirsch}, {Sinukoff}, {Cumming},
  {Hebb}, \& {Cargile}}]{2018AJ....155...48W}
{Weiss}, L.~M., {Marcy}, G.~W., {Petigura}, E.~A., {et~al.} 2018, \aj, 155, 48

\bibitem[{{Williams} {et~al.}(2019){Williams}, {Cieza}, {Hales}, {Ansdell},
  {Ruiz-Rodriguez}, {Casassus}, {Perez}, \& {Zurlo}}]{2019ApJ...875L...9W}
{Williams}, J.~P., {Cieza}, L., {Hales}, A., {et~al.} 2019, \apj, 875, L9

\bibitem[{{Wolfgang} {et~al.}(2016){Wolfgang}, {Rogers}, \&
  {Ford}}]{2016ApJ...825...19W}
{Wolfgang}, A., {Rogers}, L.~A., \& {Ford}, E.~B. 2016, \apj, 825, 19

\bibitem[{{Wu}(2019)}]{2019ApJ...874...91W}
{Wu}, Y. 2019, \apj, 874, 91

\bibitem[{Youdin \& Lithwick(2007)}]{2007Icar..192..588Y}
Youdin, A. \& Lithwick, Y. 2007, Icarus, 192, p. 588

\bibitem[{{Zeng} {et~al.}(2019){Zeng}, {Jacobsen}, {Sasselov}, {Petaev},
  {Vanderburg}, {Lopez-Morales}, {Perez-Mercader}, {Mattsson}, {Li}, \&
  {Heising}}]{2019arXiv190604253Z}
{Zeng}, L., {Jacobsen}, S.~B., {Sasselov}, D.~D., {et~al.} 2019, arXiv
  e-prints, arXiv:1906.04253

\bibitem[{{Zhang} {et~al.}(2018){Zhang}, {Zhu}, {Huang}, {Guzm{\'a}n},
  {Andrews}, {Birnstiel}, {Dullemond}, {Carpenter}, {Isella}, {P{\'e}rez},
  {Benisty}, {Wilner}, {Baruteau}, {Bai}, \& {Ricci}}]{2018ApJ...869L..47Z}
{Zhang}, S., {Zhu}, Z., {Huang}, J., {et~al.} 2018, \apjl, 869, L47

\bibitem[{{Zhu} \& {Wu}(2018)}]{2018arXiv180502660Z}
{Zhu}, W. \& {Wu}, Y. 2018, AJ, 156, id.92

\bibitem[{{Zhu} {et~al.}(2019){Zhu}, {Zhang}, {Jiang}, {Kataoka}, {Birnstiel},
  {Dullemond}, {Andrews}, {Huang}, {Perez}, {Carpenter}, {Bai}, {Wilner}, \&
  {Ricci}}]{2019arXiv190402127Z}
{Zhu}, Z., {Zhang}, S., {Jiang}, Y.-F., {et~al.} 2019, arXiv e-prints,
  arXiv:1904.02127

\end{thebibliography}
\end{document}